# Automatic Timing-Coherent Transactor Generation for Mixed-level Simulations

Li-Chun Chen

Dept. of Electrical Engineering,
National Tsing-Hua University,
Hsinchu, Taiwan
s9861524@m98.nthu.edu.tw

Hsin-I Wu

Dept. of Computer Science,
National Tsing-Hua University,
Hsinchu, Taiwan
hiwu.dery@gmail.com

Ren-Song Tsay

Dept. of Computer Science,
National Tsing-Hua University,
Hsinchu, Taiwan
rstsay@cs.nthu.edu.tw

## ABSTRACT

In this paper we extend the concept of the traditional *transactor*, which focuses on correct content transfer, to a new timing-coherent *transactor* that also accurately aligns the timing of each transaction boundary so that designers can perform precise concurrent system behavior analysis in mixed-abstraction-level system simulations which are essential to increasingly complex system designs. To streamline the process, we also developed an automatic approach for timing-coherent *transactor* generation. Our approach is actually applied in mixed-level simulations and the results show that it achieves 100% timing accuracy while the conventional approach produces results of 25% to 44% error rate.

## 1. INTRODUCTION

As system complexity continues to increase, system-level abstractions and mixed-level simulations for both functional and timing verifications are critical for design productivity. In order to reduce verification complexity, designers often focus only on one specific component at a time and keep the rest of the target system abstract model in a top-down design process [1-2]. Hence, it is crucial to have a methodology that can seamlessly integrate models of different abstraction levels while ensuring functional correctness and timing coherence for mixed-level simulations.

*Transactor* [6-11] techniques have been widely applied to enable system simulations and integrations. A transactor translates information of different formats from one component model to another. Previous works focus on the modeling and automation of *transactors* to convert the contents of the transaction but do not consider timing effects. Therefore, these approaches cannot guarantee timing coherence, for which both components connected by a *transactor* should have the same transaction timing reference. For instance, both should see the same transaction start and end time. Without timing coherence, system components cannot be correctly synchronized. Hence, system concurrent behaviors cannot be accurately captured or verified with these approaches.

To illustrate the coherence issue, we show in Figure 1(a) a model of a conventional *transactor*. We assume that a Cycle-Accurate (CA) model is on one side of the *transactor* and a Programmer-View-with-Time (PVT) model is on the other. In the CA model, all signals are examined cycle by cycle following the bus protocol, but in the PVT model, payload (i.e. the collection of bus addresses and data transferred) is passed as a parameter through a function call. For a transaction from the PVT to the CA, the *transactor* reads the PVT payload and distributes the corresponding signals to the CA model. Conversely, for a transaction from the CA to the PVT, the *transactor* collects the address and data from the CA model and packs them into a payload for the PVT model. As a result, the contents of the transaction are correctly translated.

However, the timing behavior is not consistent when translating a transaction from a CA model to a PVT model. Figure 1(b) shows the time span of a CA-to-PVT write transaction translated by a conventional *transactor*. In this case, the *transactor* can issue the *write* transaction call only after completing the payload collection, i.e., at the time point indicated by the upward arrow. Now, a subtle issue emerges: when collecting payload, the *transactor* must advance the clock cycle-by-cycle in order to process address and data transfers according to the CA interface protocol. Therefore, the conventional *transactor* approaches have to delay issuing the transaction to the PVT side until the collection process is completed. As a result, the two models each owns different time reference for the transaction and this results in compromised simulation results.

The reason why the above timing incoherent issue occurs is because a *transactor* cannot instantly determine the payload when a CA transaction is being processed. Additionally, since both the CA and PVT interface protocols are timing sensitive, any clock cycle misalignment will result in timing incoherent results.

A part of the above issue can easily be resolved by having a separate local clock for each component model on one side of the transactor to avoid any delayed content issuing. Specifically, we let the *transactor* manage the local clocks of the connected models and have each component model advance its own local clock separately. However, to fully resolve the issue we need to further develop a full timing coherence policy for the *transactor* to integrate component models of different abstraction levels while following the interface protocols. Details of our proposed timing coherence approach will be elaborated in Section 3.

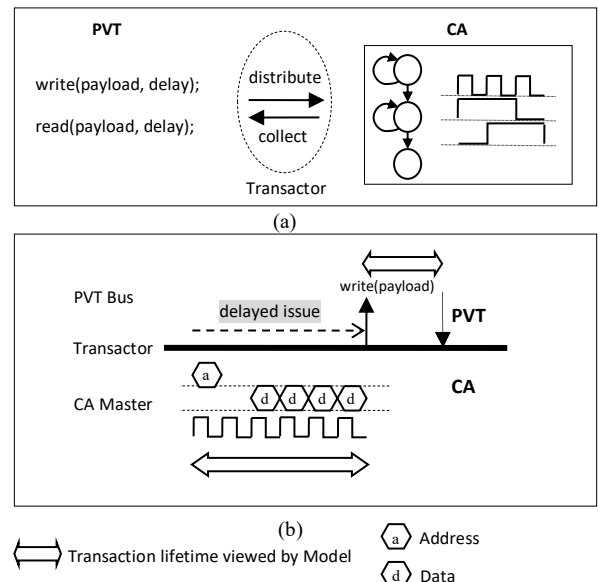

Figure 1: (a) A *transactor* connecting a PVT and a CA models. (b) Timing incoherence occurs when data is transferred from a CA model to a PVT model through a conventional *transactor*.

To alleviate designers from tedious and error-prone manual implementation, we use FSM (Finite State Machine) to describe the interface behavior of each abstraction level and devise a generation algorithm. We automatically extract timing-coherent constraints from the interface FMSs to guide our *transactor* generation algorithm and guarantee consistent beginning and ending points of every transaction. We will illustrate the details in Section 4.

Our idea has been implemented and tested on industrial cases and the experimental results show that our approach can simulate accurately while the conventional approach has a 25% to 44% of accuracy degradation.

The remainder of this paper is organized as follows: Section 2 describes related work, Section 3 explains issues in timing coherence of mixed-level simulation and the proposed transactor generation algorithm is described in Section 4, Section 5 discusses experimental results and Section 6 presents our conclusion.

## 2. RELATED WORK

*Transactors* have been widely adopted in today's SoC (System-on-Chip) design processes. They were originally used in Transaction Based Verification (TBV) [9]. By separating the verification process of an RTL IP (Register-Transfer-Level Intellectual Property) into *transactor* modeling and testbench design, a complex test suite can be built effectively. The concept is further applied to the verification of TLM (Transaction Level Modeling) models using RTL testbenches and the co-simulation of TLM and RTL models [10, 11]. These works mainly aim to translate the transaction contents between component models in different abstraction levels, and omit timing coherence consideration.

*Transactors* are also used for communication architecture exploration and design [16-18]. A TLM model can be connected to a CA bus through a *transactor* as shown in Figure 1(a). Since timing coherence is not considered, the mix-level simulation results of the architecture performance estimations are inaccurate as discussed in the introduction section.

To improve the tedious and error-prone *transactor* modeling process, researchers have also proposed a few automatic *transactor* generation algorithms [6-8]. However, none of these approaches address the timing coherence problem in connecting component models of different abstraction levels. Therefore, they are only suitable for functional verification rather than performance estimation or concurrent behavior simulations.

Despite not being able to guarantee timing coherence, Bombieri et. al. [6] propose a template-based approach for *transactor* generation. The approach is restricted to AHB (AMBA High-performance Bus)-like protocol and cannot handle pipelined and out-of-order bus protocols. Balarin et al. in [7] propose a finite automaton-based approach that is widely adopted for *converter* (or *transducer*) synthesis [19-20] with protocol behavior captured by one or multiple FSMs (Finite State Machines) depending on the concurrency property. We adopt and extend the approach in [20] for our purpose as out-of-order protocol behavior can be captured easily by multiple FSMs.

The study of *transactor* generation has also attracted industrial attention. Examples are *TransactorWizard* from Structured Design Verification [12], *BusCompiler* from Synopsys [13] and *Cohesive* from Spiratech [14]. However, due to the proprietary nature of these commercial tools, we omit making direct comparisons with these tools.

As aforementioned, none of these approaches can achieve timing-coherent mixed-level simulations. Our proposal is the first automatic *transactor* generation approach that can guarantee timing-coherent results of mixed-level system simulation. To explicate the idea, timing coherence issue is first elaborated.

## 3. THE PROPOSED APPROACH

Without loss of generality, in the following we use the example of connecting a CA component model to a PVT component model to explain our approach.

First, note that both payload content correctness and timing coherence are required for performing mixed-level simulations. To ensure payload correctness, the mapping between signals in the CA (i.e. address and data) and payload in the PVT has to be encoded in the *transactor*, which distributes address and data or collects payload according to the given protocols. To guarantee timing coherence, as discussed in the introduction, we have a separated local clock for each component model and then synchronize timing only at the transaction boundaries. Before describing the details of our approach, we will first present the timing coherence issues in the TLM abstraction levels.

### 3.1 Timing coherence

As shown in Figure 2, components modeled at different abstraction levels may have different timing details of a transaction. A PVT model concerns only the beginning and ending points of the transaction. A Bus-Accurate (BA) model, however, also requires the beginning and ending time points of each sub-transaction, such as the request phase or data phase defined in OCP [21]. A CA model defines the behavior of every cycle.

To simplify our later discussion, in Figure 2 we use white dots and black dots to indicate the beginning and ending points of transactions (or sub-transactions) respectively.

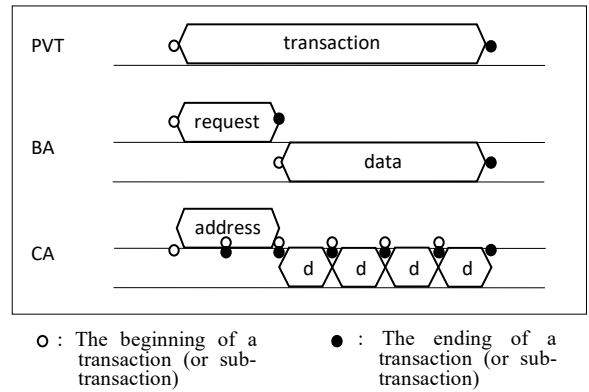

Figure 2: An illustration of timing coherence among models in different TLM abstraction levels.

Now we say that a *transactor* is timing-coherent if the communications between the two connected component models not only have correct functionality but also exhibit consistent transaction beginning and ending timing points.

## 3.2 Local clock management

To construct a timing-coherent *transactor*, the key is to have the *transactor* manage the local clocks of the connected component models. A *transactor* should separately advance the clock of each connected component model and coordinate the clocks to ensure coherent global timing.

Essentially, instead of having the simulators control clock signals as in the conventional approaches, we let the *transactor* control the local clocks as shown in Figure 3(a). In this scheme, before a payload is issued, the *transactor* advances only the local clock of the low level component model while the clock of the high level component model is not affected. Once the payload information is fully collected, then the *transactor* issues the transaction call and advances the local clock of the high level component model to the correct transaction ending time. We name this process *local clock wrapping* as the *transactor* wraps the local clock behavior of the low level model around the transactional information for a higher level model.

The timing diagrams in Figures 3(b)~3(e) illustrate the local clock wrapping process. We assume that the CA master component model is below the solid line while the PVT bus component model is above the solid line, and the solid line itself represents the *transactor*. As the CA master issues a transfer, the *transactor* records the beginning time of the transfer and drives the local clock until it finishes collecting the payload. Note that we drive only the *local clock* of the CA component as shown in Figure 3(b), and that the time of the CA model is now 50 ns while the time of the PVT model remains at 0 ns (assuming the clock period is 10 ns). At this point, the content of the payload is valid and the *transactor* issues a transaction call to the PVT component model as shown in Figure 3(c). Assuming that the returned delay is 50 ns, the *transactor* then advances the clock of the PVT model, as shown in Figure 3(d), so that the beginning and the ending points of the transaction are consistent on both sides of the *transactor* as shown in Figure 3(e).

In the following we shall discuss the case that a transaction is only determinable at runtime.

## 3.3 Dynamic timing matching

Models of different abstraction levels not only have different timing view of a transaction but also have different ways of managing delay behaviors. For instance, a PVT model simply takes a delay parameter while a CA model adopts a signal handshake to determine the actual delay values.

The examples in Figure 3(f)-(g) illustrate the scenario that the CA model recognizes the end of a transaction when the handshake signal of the last data transfer is raised (such as HREADY signal in the AHB bus protocol). In this case, the returned delay from the PVT bus is not the ideal 50 ns value. We assume that the returned delay is 70 ns, which includes an additional contention delay. To be consistent, our proposed approach automatically detects from the interface protocol and lets the *transactor* keep the HREADY signal low for an additional 20 ns so that the transaction end points on both sides would match exactly. Details are discussed in the next section.

Although the *transactor* mechanism proposed above is effective, it would be tedious and error-prone if implemented manually. In order to relieve designers from this tedious process, we devise an algorithm for automatic timing-coherent *transactor* generation as discussed in the following.

## 4. AUTOMATIC GENERATION

We propose a FSM-based generation algorithm that can automatically generate a timing-coherent *transactor* from the two given FSMs that describe the interface behavior of the connected components. Essentially, our algorithm generates a third FSM that represents the *transactor* to interact between the given FSMs and to conduct transactions with both correct function and coherent timing.

## 4.1 Interface FSM specification

First, we define the notations used in the FSM that describes the behavior of an interface protocol. As shown in Figure 4, the notation '!' indicates the setting of signals and '?' represents the reading of signals. We name the FSM issuing a transaction as the *Initiator* and the one responding to the transaction as the *Target*. Note that the *Initiator* and *Target* complement each other and both can progress synchronously to accomplish a transaction.

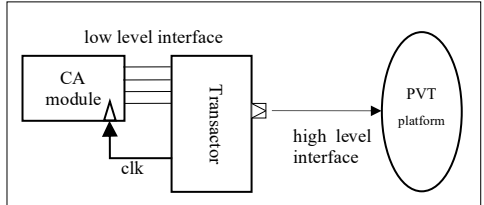

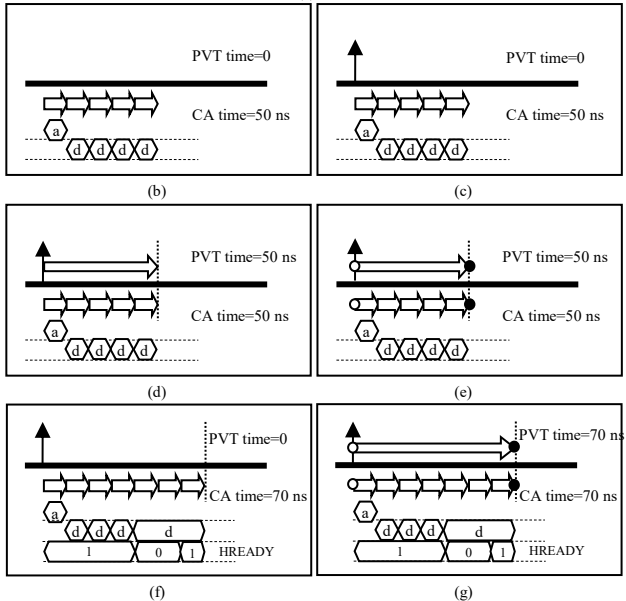

Figure 3: (a) An overview of the transactor architecture. (b) The *transactor* collects payload from the CA-level model component and advances its clock. (c) The *transactor* issues the collected payload to the PVT model component at the right time point. (d) The *transactor* advances the PVT clock to match the ending point of the transaction. (e) The *transactor* sets consistent transaction boundaries. (f) The *transactor* keeps the HREADY signal low for 20 ns more. (g) The timing coherent result for the case in (f).

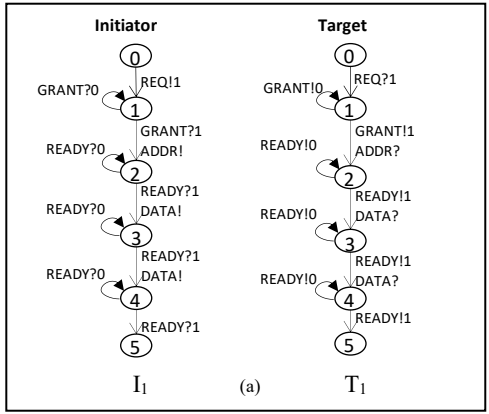

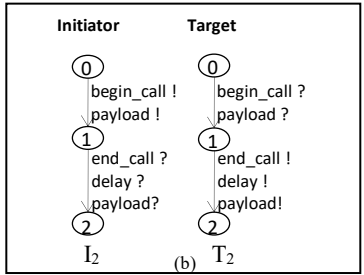

Figure 4: (a) An example FSM of a CA interface pair; (b) A sample FSM representation of a transaction-level interface pair.

For example, Figure 4(a) shows a complementary CA-level interface pair that describes the behavior of transferring one address and two data via cycle-by-cycle signal handshaking. Conversely, Figure 4(b) shows a complementary FSM pair for a PVT, which is a higher transaction-level interface model whose transition is associated with a function call and parameter passing; there is no clock triggering and no signal setting, unlike the CA model. Note that in a higher-level TLM environment, the communication is mostly done through IMC (Interface Method Call) instead of signal handshaking.

We now examine the PVT *Initiator* FSM in Figure 4(b) in more detail. In the transition from state 0 to state 1, a transaction call is initiated (begin_call!) and the payload is sent to the *Target* (payload!). Then, to progress the transition from state 1 to state 2, we have to wait for the notification for the end of the transaction call along with a returned transaction delay value, and the response field of the payload being filled in.

Since a transaction always occurs between two components, we take the Initiator FSM of one component and the Target FSM of the other component as the input for our *transactor* generation algorithm.

## 4.2 Transactor generation

There are two steps to our generation process as shown in Figure 5. The first step is to find the complementary form of the two input FSMs of the components to be connected. Second, our generation algorithm explores the two complementary FSMs and generates a timing-coherent *transactor*.

### 4.2.1 Complementary FSM

To connect components of different abstraction levels, the *transactor* must be able to communicate with the FSMs that represent the interface behaviors of the two components. Thus, the

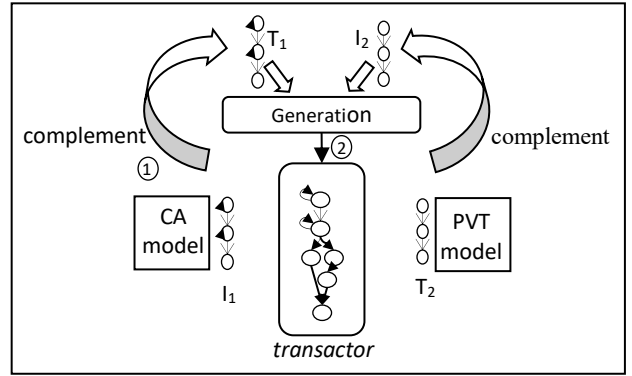

Figure 5: An overview of the generation process

*transactor* to be constructed is simply a composition of the two complementary FSMs of the two original FSMs of interest. The complement of an FSM is simply the inversion of all its actions. For instance, the *Initiator* and *Target* shown in Figure 4, each is a complement to the other.

If we assume that $I_1$ is the initiator FSM of the first component and $T_1$ is its complementary FSM. Correspondingly, $T_2$ is the input FSM of the second component and $I_2$ is its complementary FSM. Then, $I_1$ and $T_2$ will be the inputs to our generation process and their complements, i.e., $I_2$ and $T_1$, will be used to synthesize the *transactor* as shown in Figure 5.

### 4.2.2 Automatic Generation Algorithm

Now, we elaborate the *transactor* generation algorithm which is extended from the synthesis approaches in [19-20]. The key of the algorithm is to properly coordinate the two complementary FSMs and make sure that all transitions of the resultant *transactor* are legal (i.e., the *transactor* translates the transaction correctly and is timing-coherent). For a transition to be *legal*, the payload has to be transferred correctly and the timing has to be coherent.

To check the legality of a transition, we first examine what qualifies as a legal transition of one state pair to the next as shown in Figure 6. Here, we use the term *state pair* to indicate a pair of states on the two complementary FSMs (e.g., $I_2$ and $T_1$ shown in the previous example) to be synthesized for the *transactor*. The two initial states of the FSMs naturally form the first state pair. Next we consider the process of traversing an intermediate state pair. We then can generalize it to the complete transition process.

We assume that there are $n$ possible transitions from one of the intermediate states, say state $P$, of $T_1$ (one of the FSM pair) and $m$ next transitions from another intermediate state $Q$ of $I_2$. The task is to determine which transitions are legal among the possible $n + m$ transitions. Any transition that violates legality is excluded.

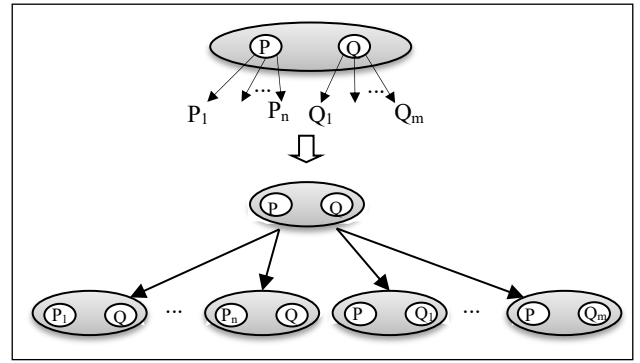

Figure 6: A typical exploration step.

For payload correctness, we have to ensure that the payload has all the required information being transferred, if the transition triggers a payload transfer. Any transition that violates the requirement is also excluded.

For timing coherence, we leverage the local clock wrapping and dynamic timing matching technique discussed in Section 3 to ensure coherence. We say that a state-pair transition is not timing-coherent if the target state reaches a transaction boundary while the initiator state still cannot determine the transaction time.

Then with local clocks being advanced separately, we choose only one transition from either state $P$ or state $Q$ instead of a concurrent transition pair at each step for progression as shown in Figure 6. Thus, for each step we only need to check at most $n + m$ transitions instead of $n \times m$ transitions. In this way, the search algorithm can perform efficiently.

Furthermore, for dynamic timing matching, the *transactor* has to translate the returned delay of the PVT model into a number of cycles in the CA model by keeping the handshake signal inactive for the corresponding period of time. To achieve this, we parse the FSM of the CA interface to find the *last handshake* state, such as state 4 in Figure 4(a), and convert it into a delay consumption model as shown in Figure 7(b). In this way, any transition to the final state that occurs before the receipt of the delay parameter is regarded as an illegal transition since there would be no delay to consume. The *transactor* essentially keeps the handshake signal inactive at this state until the specified delay is consumed. By doing so, the ending point of the transaction on the PVT model according to the delay parameter can be matched with that of CA at the last transition.

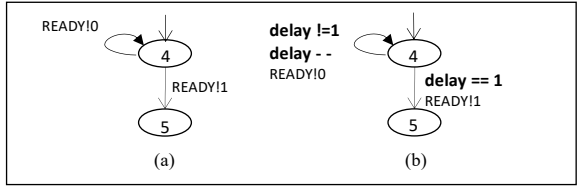

Figure 7: (a) The last handshake state of the *Target* in Figure 4(a). (b) Adding a delay consumption model.

Moreover, we need to do special processing on control signals, such as the READY signal, which occur in self-loops and affect timing as shown in Figure 4(a). For our approach, we simply remove these self-looping transitions except the last one before completion of *transactor* generation and adjust the timing at the last handshake state as discussed,

The generation algorithm is summarized below for reference. We apply the DFS (Depth-First-Search) to explore the combined states of the input FSM pair. In addition to the FSM pair, the mapping of the payload, such as L={addr, data, data} for the example in Figure 4, is assumed to be provided by designers.

The two initial states of the input FSMs naturally form the first state pair which will be the first to be explored. At each step, the algorithm checks the consistency of the data and timing of all the out-going transition to ensure payload correctness and timing coherence. Any transition that violates the constraints will not be explored further. We also check if the transition is a self-loop and skip it to avoid infinite recursion. If the transition is legal, then we add it into the output *transactor* FSM. If the state pair being visited is the final state pair, we directly add it into the output FSM and exit and determine that a legal *transactor* is found. If a state pair has already been visited, then we know that all its descendants have been explored and no transitions of this state pair can reach final state pair. We then remove this state and continue exploring other states until a legal *transactor* found or exhausted.

**Algorithm:** Generation(T, I, L)

1. Input: A FSM pair T,I ; The mapping of payload L
2. Output: A transactor FSM G, initialized to Ø
3. S: A stack of {p, q} = Ø for DFS
4. **Begin**
5.   S.push(<0,0>); // push initial state
6.   **while** S != Ø **do**
7.   **begin**
8.     {p, q} = S.pop();  // p is a state in T and q is a state in I
9.     **if** {p, q} has been visited, G.delete( all outgoing transitions from {p,q} ) and **go to** step 6;
10.     mark {p,q} visited;
11.     **if** leaf-node, **exit**; //success
12.     **for** each outgoing transition p→ p'
13.       **if**( data inconsistent || timing inconsistent ) **continue**; //to next transition
14.       **if**( p' != p) S.push( {p', q} ); // do not follow SELF-LOOP;
15.       G.add( {p,q}→{p',q} );
16.     **end-for**
17.     **for** each outgoing transition q→ q'
18.       **if**( data inconsistent || timing inconsistent ) **continue**; //to next transition
19.       **if**( q'! = q) S.push( {p, q'} );// do not follow SELF-LOOP;
20.       G.add( {p,q}→{p,q'} );
21.     **end-for**
22.   **end-while**
23. **End**

In Figure 8, we show the resultant *transactor* FSM of the input *Initiator* in Figure 4(a) and the *Target* in Figure 4(b). This *transactor* can correctly perform timing-coherent translation as shown in Figure 3.

## 5. CASE STUDY

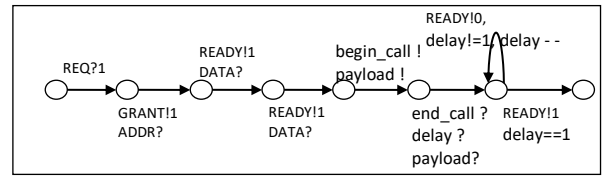

Figure 8: The result of the transactor generated from the *Initiator* in Figure 4(a) and *Target* in Figure 4(b).

We have implemented our proposed timing-coherent *transactor* generation algorithm and verified its effectiveness on a few industrial designs. The experiments were performed on a platform equipped with an Intel Xeon 3.4GHz quad-core and 2GB ram. The target design was modeled in SystemC 2.2.0.

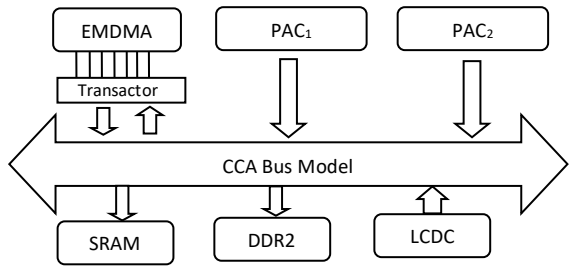

Figure 9: PAC Duo platform

We took the PAC platform [18] shown in Figure 9 for a case study. For the experiment, we adopted a CCA (Cycle-Count Accurate) bus model [5], along with the rest of the PVT models except for the EMDMA. The EMDMA is of the CA abstraction model with a CoWare communication interface. We connected the EMDMA to the high level platform through a *transactor*. We compared our timing-coherent approach with the traditional integration approach. We computed the accuracy rate of each transaction by comparing the beginning and ending times of the transaction. We found that our approach can accurately capture the timing of each transaction while the conventional approach has a 25% to 44% inaccuracy rate.

In Table 1, we list the testing results of three testbenches for the EMDMA module in Figure 9. The *general channel test* contains burst transfers of various lengths and the *multimedia test* includes common context tasks in multimedia applications (which involve transferring data from two sources to one destination or from one source to two destinations). The last column shows the error rate of each transaction from executing the H264 code. With the local clock wrapping and dynamic timing matching, our approach achieves 100% accuracy on the full system simulation.

Table 1: The accuracy rate comparison

|  | General Channel Test | Multimedia Test | H264 |
|---|---|---|---|
| Error Rate of Our Approach | 0% | 0% | 0% |
| Error Rate of the Conventional Approach | 44% | 25% | 42% |

## 6. CONCLUSION

This paper proposes an automatic *transactor* generation approach that can guarantee timing-coherent results in mixed-level system simulations. We demonstrate that our approach achieves 100% timing accuracy in mixed level system simulations while the conventional approach produces a 25% to 44% error rate. Our proposed approach ensures reliable performance estimation and accurate system concurrent behavior simulations. The method can be generalized to perform system integration of high-abstraction software models with hardware implemented models.